\documentclass[11pt]{article}
\usepackage{amssymb,amsmath,amsfonts}
\usepackage{graphicx}
\usepackage{graphics}
\usepackage{eepic,epsfig}

\@addtoreset{equation}{section}

\makeatother

\begin{document}

\title{Fermionic Casimir densities in anti-de Sitter spacetime}
\author{E. Elizalde$^{1}$\thanks{%
E-mail: elizalde@ieec.uab.es},\thinspace\ S.~D. Odintsov$^{1,2}$\thanks{%
E-mail: odintsov@ieec.uab.es; also at Tomsk State Pedagogical University
(Tomsk) and Eurasian National University (Astana)}, \thinspace\ A.~A.
Saharian$^{3}$\thanks{%
E-mail: saharian@ysu.am} \\
\\
\textit{$^{1}$Instituto de Ciencias del Espacio (CSIC) }\\
\textit{and Institut d'Estudis Espacials de Catalunya (IEEC/CSIC) }\\
\textit{Campus UAB, Facultat de Ci\`{e}ncies, Torre C5-Parell-2a planta,}\\
\textit{08193 Bellaterra (Barcelona) Spain}\vspace{0.3cm}\\
\textit{$^2$Instituci\'{o} Catalana de Recerca i Estudis Avan\c{c}ats (ICREA)%
}\vspace{0.3cm}\\
\textit{$^3$Department of Physics, Yerevan State University,}\\
\textit{1 Alex Manoogian Street, 0025 Yerevan, Armenia}}
\maketitle

\begin{abstract}
The fermionic condensate and vacuum expectation value of the
energy-momentum tensor, for a massive fermionic field on the
background of anti-de Sitter spacetime, in the geometry of two
parallel boundaries with bag boundary conditions, are
investigated. Vacuum expectation values, expressed as series
involving the eigenvalues of the radial quantum number, are neatly
decomposed into boundary-free, single-boundary-induced, and
second-boundary-induced parts, with the help of the generalized
Abel-Plana summation formula. In this way, the renormalization
procedure is very conveniently reduced to the one corresponding to
boundary-free AdS spacetime. The boundary-induced contributions to
the fermionic condensate and to the vacuum expectation value of
the energy density are proven to be everywhere negative. The
vacuum expectation values are exponentially suppressed at
distances from the boundaries much larger than the curvature
radius of the AdS space. Near the boundaries, effects related with
the curvature of the background spacetime are shown to be
subdominant and, to leading order, all known results for
boundaries in the Minkowski bulk are recovered. Zeta function
techniques are successfully used for the evaluation of the total
vacuum energy in the region between the boundaries. It is proven
that the resulting interaction forces between them are attractive
and that, for large separations, they also decay exponentially.
Finally, our results are extended and explicitly translated to
fermionic Casimir densities in braneworld scenarios of
Randall-Sundrum type.
\end{abstract}

\bigskip

\section{Introduction}

Anti-de Sitter (AdS) spacetime is among the most popular background
geometries in quantum field theory. Much of the earlier interest in this
geometry was motivated by questions of principal nature, mainly related with
the quantization of fields on curved backgrounds. The AdS spacetime has
maximal symmetry and, because of this, numerous physical problems can be
exactly solved on this background. The presence of both regular and
irregular modes and the possibility of getting an interesting causal
structure lead to several new and remarkable phenomena. Further interest in
this subject arose from the discovery that the AdS spacetime generically
arises as a ground state in extended supergravity and string theories, what
is again potentially most important.

In recent developments of the topic, the AdS geometry is an arena for two
classes of models. The first is the AdS/CFT correspondence (for a review see
\cite{Ahar00}), which represents a realization of the holographic principle
and relates string theories or supergravity in the AdS bulk with a conformal
field theory living on its boundary. This correspondence has many
interesting consequences and provides a powerful tool for the investigation
of physical effects in gauge theories. The second class of models with the
AdS spacetime as background geometry is a realization of the braneworld
scenario with large extra dimensions and provides a solution to the
hierarchy problem which arises between the gravitational and electroweak
mass scales (for reviews on braneworld gravity and cosmology see \cite%
{Brax03}). Here the small coupling of 4-dimensional gravity is generated by
the large physical volume of extra dimensions. Braneworlds naturally appear
in the string/M theory context and provide a novel set up for the discussion
of phenomenological and cosmological issues related with extra dimensions.

In the present paper, as an example of an exactly solvable physical problem
in AdS spacetime, we will consider the Casimir effect (for reviews see \cite%
{Casimir}) for a fermionic field obeying bag boundary conditions on two
parallel plates. The explicit dependence of the characteristics of the
vacuum on the geometry of the background spacetime is among the most
interesting topics in the investigation of the Casimir effect. As usual, all
relevant information is encoded in the spectrum of the vacuum fluctuations
and, not surprisingly, analytic solutions can only be found, in general, for
highly symmetric geometries. Specifically, the Casimir effect for a massive
scalar field with general curvature coupling parameter in the geometry of
flat and spherical boundaries on the background of de Sitter spacetime has
been investigated recently in \cite{Saha09} and \cite{Milt12}, respectively.

Investigations of the Casimir effect in AdS spacetime have already attracted
a great deal of attention, motivated by Randall-Sundrum type braneworld
scenarios \cite{Rand99}. In these models the background solution consists of
two parallel flat branes embedded in a 5-dimensional AdS bulk. The fifth
coordinate is compactified on $S^{1}/Z^{2}$, and the branes are on two fixed
points. The fields which propagate in the bulk will give Casimir-type
contributions to the vacuum energy and, as a result, to the vacuum forces
acting on the branes. The Casimir effect provides in this context a natural
mechanism for stabilizing the radion field, as required for a complete
solution of the hierarchy problem. In addition, the Casimir energy gives a
contribution to both the brane and the bulk cosmological constants and,
hence, it has to be taken into account in any self-consistent formulation of
the braneworld dynamics. The Casimir energy and the corresponding forces for
two parallel branes in AdS spacetime have been evaluated in Refs.~\cite%
{Gold00,Flac01}, both for scalar and fermionic fields, by using either
dimensional or zeta function regularization methods. Local Casimir densities
were considered in Refs.~\cite{Knap04,Saha05,Shao10}. The Casimir effect in
higher-dimensional generalizations of the AdS spacetime with compact
internal spaces has been investigated in~\cite{Flac03} while the Casimir
energy for a massless fermionic field with generalized bag boundary
conditions in 3-dimensional AdS spacetime was discussed in \cite{Bene10},
for the geometry where one of the boundaries coincides with the AdS boundary.

In braneworld models, two distinct types of boundary conditions arise for
Dirac fermion fields, corresponding respectively to even and odd fields
(untwisted and twisted boundary conditions) \cite{Gher00}. In the present
paper, we investigate one-loop quantum effects for a fermionic field in AdS
spacetime, induced by two parallel boundaries with bag boundary condition.
Although this condition is different from those appearing in braneworld
scenarios, we will here show in detail how, from our formulas, the
corresponding results for braneworlds are readily obtained. The important
quantities that characterize the local properties of the fermionic vacuum
are the fermionic condensate (FC) and the vacuum expectation value (VEV) of
the energy-momentum tensor. For the investigation of these quantities we use
the direct mode summation approach, which requires the knowledge of a
complete set of mode functions for the fermionic field obeying the boundary
conditions.

In the next section, we describe the geometry of the problem and construct
the corresponding mode functions. By making use of them, in Sect.~\ref%
{sec:FC} we evaluate the FC. Applying the generalized Abel-Plana formula,
the FC is decomposed into three parts: a boundary-free contribution, a
single-boundary-induced one, and a second-boundary-induced one. The behavior
of the VEVs in asymptotic regions of the parameters is discussed. In Sect.~%
\ref{sec:EMT} we present similar considerations for the VEV of the
energy-momentum tensor. Casimir forces acting on the boundaries and the
corresponding Casimir energy are investigated in Sect.~\ref{sec:Force}. The
generalization of the formulas obtained to the important case of a bulk
fermionic field in the Randall-Sundrum braneworld model is discussed.
Finally, our main results are summarized in Sect.~\ref{sec:Conc}.

\section{Fermionic mode functions}

\label{sec:Modes}

We consider a quantum fermionic field, $\psi $, in $(D+1)$-dimensional
anti-de Sitter spacetime, $AdS_{D+1}$. In Poincar\'{e} coordinates, the
corresponding line element reads
\begin{equation}
ds^{2}=g_{\mu \nu }dx^{\mu }dx^{\nu }=e^{-2y/a}\eta _{ik}dx^{i}dx^{k}-dy^{2},
\label{ds2deSit}
\end{equation}%
with $\eta _{ik}=\mathrm{diag}(1,-1,\ldots ,-1)$, $i,k=0,\ldots ,D-1$, being
the Minkowskian metric tensor for a $D$-dimensional spacetime. In (\ref%
{ds2deSit}), $a$ is the AdS curvature radius which is related to the Ricci
scalar by $R=-D(D+1)/a^{2}$. In addition to the radial coordinate $y$ we
will use the conformal coordinate $z$, defined as $z=ae^{y/a}$. With this
coordinate, the AdS line element is written in conformally-flat form:%
\begin{equation}
ds^{2}=(a/z)^{2}\left( \eta _{ik}dx^{i}dx^{k}-dz^{2}\right) .
\label{ds2Conf}
\end{equation}%
The hypersurfaces $z=0$ and $z=\infty $ correspond to the AdS boundary and
to the horizon, respectively.

The dynamics of a fermionic field in curved spacetime are governed by the
covariant Dirac equation
\begin{equation}
i\gamma ^{\mu }\nabla _{\mu }\psi -m\psi =0\ ,\;\nabla _{\mu }=\partial
_{\mu }+\Gamma _{\mu },  \label{Direq}
\end{equation}%
where $\Gamma _{\mu }$ is the spin connection. The Dirac matrices $\gamma
^{\mu }$ are expressed in terms of the flat spacetime gamma matrices $\gamma
^{(a)}$, as $\gamma ^{\mu }=e_{(a)}^{\mu }\gamma ^{(a)}$, with $e_{(a)}^{\mu
} $ being the tetrad fields obeying the relation $e_{(a)}^{\mu }e_{(b)}^{\nu
}\eta ^{ab}=g^{\mu \nu }$. For the spin connection, one has
\begin{equation}
\Gamma _{\mu }=\frac{1}{4}\gamma ^{(a)}\gamma ^{(b)}e_{(a)}^{\nu }e_{(b)\nu
;\mu }\ ,  \label{Gammamu}
\end{equation}%
where the semicolon means covariant derivative of vector fields. For the
geometry under consideration the tetrads can be taken in the form $%
e_{(b)}^{\mu }=\delta _{b}^{\mu }z/a$. With this choice, the spin connection
has the following components (no summation over $l$)%
\begin{equation}
\Gamma _{D}=0,\;\Gamma _{l}=\frac{\eta _{ll}}{2z}\gamma ^{(D)}\gamma
^{(l)},\;l=0,\ldots ,D-1.  \label{SpinCon}
\end{equation}%
For the combination appearing in the Dirac equation, we have $\gamma ^{\mu
}\Gamma _{\mu }=-D\gamma ^{(D)}/(2a)$.

We are interested in the change of the properties of the fermionic vacuum
induced by the presence of the two boundaries, which are located at $z=z_{1}$
and $z=z_{2}$, $z_{1}<z_{2}$. The corresponding values of the physical
radial coordinate, $y$, will be denoted by $y_{1}$ and $y_{2}$: $%
z_{j}=ae^{y_{j}/a}$, $j=1,2$. We will assume that, on the boundaries, the
field obeys the bag boundary condition, namely
\begin{equation}
(1+i\gamma ^{\mu }n_{\mu }^{(j)})\psi =0,\;z=z_{j},  \label{Bagbc}
\end{equation}%
$j=1,2$, with $n_{\mu }^{(j)}$ being normal to the boundaries, $n_{\mu
}=(-1)^{j}\delta _{\mu }^{D}a/z$. From these conditions it follows that the
normal component of the fermion current vanishes at the boundaries.

In $(D+1)$-dimensional flat spacetime the Dirac matrices are $N\times N$
matrices, with $N=2^{[(D+1)/2]}$ (the square brackets mean the integer part
of the enclosed expression). In the discussion below we will assume the
following representation for these matrices:%
\begin{equation}
\gamma ^{(0)}=i\left(
\begin{array}{cc}
0 & -1 \\
1 & 0%
\end{array}%
\right) ,\;\gamma ^{(a)}=i\left(
\begin{array}{cc}
-\sigma _{a} & 0 \\
0 & \sigma _{a}%
\end{array}%
\right) ,  \label{gamflat}
\end{equation}%
with $a=1,\ldots ,D$ and $\sigma _{a}\sigma _{b}+\sigma _{b}\sigma
_{a}=2\delta _{ab}$. The last relation directly follows from the
anticommutation relations for the Dirac matrices. The matrices (\ref{gamflat}%
) are related to the gamma matrices in the standard Dirac representation, $%
\gamma _{\mathrm{(D)}}^{(a)}$, by $\gamma ^{(a)}=iB\gamma _{\mathrm{(D)}%
}^{(a)}$, where%
\begin{equation}
B=\left(
\begin{array}{cc}
0 & 1 \\
1 & 0%
\end{array}%
\right) .  \label{B}
\end{equation}%
As we will see below, with the representation (\ref{gamflat}) the equations
for the components of the fermionic field are conveniently separated. This
fact is to be remarked, since it allows for a complete calculation.

Among the most important characteristics of the fermionic vacuum are the FC
and the VEV of the energy-momentum tensor. For the evaluation of these
quantities we will use the direct mode summation approach. In this approach
we need a complete set of solutions to Eq.~(\ref{Direq}) obeying the
boundary conditions (\ref{Bagbc}). For the positive-energy solutions the
dependence of the mode functions on the time and on the coordinates parallel
to the boundaries, denoted as $\mathbf{x}=(x^{1},\ldots ,x^{D-1})$, can be
expressed in the form $e^{i\mathbf{kx}-i\omega t}$, where $\mathbf{k}%
=(k_{1},\ldots ,k_{D-1})$, $\mathbf{kx}=k_{l}x^{l}$, and the summation runs
over $l=1,\ldots ,D-1$.

Decomposing the field into upper and lower components,%
\begin{equation}
\psi =\left(
\begin{array}{c}
\psi _{+}(z) \\
\psi _{-}(z)%
\end{array}%
\right) e^{i\mathbf{kx}-i\omega t},  \label{decomp1}
\end{equation}%
the initial Dirac equation is reduced to the set%
\begin{equation}
\left[ \sigma _{D}\left( \partial _{z}-\frac{D}{2z}\right) +ik_{l}\sigma
_{l}\mp \frac{ma}{z}\right] \psi _{\pm }-i\omega \psi _{\mp }=0,
\label{psipm}
\end{equation}%
from where we can obtain separate equations for the upper and lower
components:%
\begin{equation}
\left( z^{2}\partial _{z}^{2}-Dz\partial _{z}+\lambda
^{2}z^{2}+D^{2}/4+D/2-m^{2}a^{2}\pm \sigma _{D}ma\right) \psi _{\pm }=0,
\label{Eqpsipm}
\end{equation}%
being $\lambda ^{2}=\omega ^{2}-k^{2}$. With the substitution
\begin{equation}
\psi _{\pm }(z)=z^{(D+1)/2}\chi _{\pm }(z),  \label{xipm}
\end{equation}%
Eq.~(\ref{Eqpsipm}) is reduced to the Bessel equation:%
\begin{equation}
\left( z^{2}\partial _{z}^{2}+z\partial _{z}+\lambda ^{2}z^{2}\pm \sigma
_{D}ma-m^{2}a^{2}-1/4\right) \chi _{\pm }=0.  \label{Eqxipm}
\end{equation}

Taking the $\sigma _{D}$ matrix in the form
\begin{equation}
\sigma _{D}=\left(
\begin{array}{cc}
1 & 0 \\
0 & -1%
\end{array}%
\right) ,  \label{sigD}
\end{equation}%
we further decompose $\chi _{\pm }(z)$ into upper and lower components,%
\begin{equation}
\chi _{\pm }(z)=\left(
\begin{array}{c}
\varphi _{\pm \uparrow }(z) \\
\varphi _{\pm \downarrow }(z)%
\end{array}%
\right) .  \label{decompxi}
\end{equation}%
The solutions for these components directly follow from (\ref{Eqpsipm}), and
are%
\begin{eqnarray}
\varphi _{\pm \uparrow } &=&C_{\pm \uparrow }^{(J)}J_{ma\mp 1/2}(\lambda
z)+C_{\pm \uparrow }^{(Y)}Y_{ma\mp 1/2}(\lambda z),  \notag \\
\varphi _{\pm \downarrow } &=&C_{\pm \downarrow }^{(J)}J_{ma\pm 1/2}(\lambda
z)+C_{\pm \downarrow }^{(Y)}Y_{ma\pm 1/2}(\lambda z).  \label{phipm}
\end{eqnarray}%
where $J_{\nu }(x)$ and $Y_{\nu }(x)$ are Bessel and Neumann functions,
respectively. The relation between the coefficients in these linear
combinations are obtained by using Eq.~(\ref{psipm}). In order to find them
we note that, from the anticommutation relations for the matrices $\sigma
_{D}$ and $\sigma _{l}$, $l=1,\ldots ,D-1$, and from (\ref{sigD}), it
readily follows that the matrices $\sigma _{l}$ have the form%
\begin{equation*}
\sigma _{l}=\left(
\begin{array}{cc}
0 & b_{l} \\
c_{l} & 0%
\end{array}%
\right) ,
\end{equation*}%
with $b_{l}c_{l}=c_{l}b_{l}=1$ and $b_{l}c_{k}=-b_{k}c_{l}$,$%
\;c_{l}b_{k}=-c_{k}b_{l}$, for$\;l\neq k$. Using these relations, from the
equation (\ref{psipm}) with the upper sign, we find%
\begin{eqnarray*}
&& \omega C_{-\uparrow }^{(J)} =k_{l}b_{l}C_{+\downarrow }^{(J)}+i\lambda
C_{+\uparrow }^{(J)},\;\omega C_{-\uparrow }^{(Y)}=k_{l}b_{l}C_{+\downarrow
}^{(Y)}+i\lambda C_{+\uparrow }^{(Y)}, \\
&& \omega C_{-\downarrow }^{(J)} =k_{l}c_{l}C_{+\uparrow }^{(J)}+i\lambda
C_{+\downarrow }^{(J)},\;\omega C_{-\downarrow
}^{(Y)}=k_{l}c_{l}C_{+\uparrow }^{(Y)}+i\lambda C_{+\downarrow }^{(Y)}.
\end{eqnarray*}

As a result, the solution of the Dirac equation can be written in the form%
\begin{equation}
\psi =\frac{z^{(D+1)/2}}{\omega }e^{i\mathbf{kx}-i\omega t}\left(
\begin{array}{c}
\omega Z_{\uparrow ,\nu -1}(\lambda z) \\
\omega Z_{\downarrow ,\nu }(\lambda z) \\
k_{l}b_{l}Z_{\downarrow ,\nu }(\lambda z)+i\lambda Z_{\uparrow ,\nu
}(\lambda z) \\
k_{l}c_{l}Z_{\uparrow ,\nu -1}(\lambda z)+i\lambda Z_{\downarrow ,\nu
-1}(\lambda z)%
\end{array}%
\right) ,  \label{psi2}
\end{equation}%
with the notations%
\begin{equation}
\nu =ma+1/2,  \label{nu}
\end{equation}%
and%
\begin{eqnarray}
Z_{\uparrow ,\mu }(\lambda z) &=&C_{+\uparrow }^{(J)}J_{\mu }(\lambda
z)+C_{+\uparrow }^{(Y)}Y_{\mu }(\lambda z),  \notag \\
Z_{\downarrow ,\mu }(\lambda z) &=&C_{+\downarrow }^{(J)}J_{\mu }(\lambda
z)+C_{+\downarrow }^{(Y)}Y_{\mu }(\lambda z).  \label{Zi}
\end{eqnarray}

We now must impose the boundary conditions (\ref{Bagbc}). From the BC at $%
z=z_{1}$, we have%
\begin{equation}
\frac{C_{+\uparrow }^{(Y)}}{C_{+\uparrow }^{(J)}}=\frac{C_{+\downarrow
}^{(Y)}}{C_{+\downarrow }^{(J)}}=-\frac{J_{\nu }(\lambda z_{1})}{Y_{\nu
}(\lambda z_{1})},  \label{Coefs}
\end{equation}%
and hence%
\begin{equation}
\frac{Z_{\uparrow ,\mu }(\lambda z)}{C_{+\uparrow }^{(J)}}=\frac{%
Z_{\downarrow ,\mu }(\lambda z)}{C_{+\downarrow }^{(J)}}=Z_{\mu }(\lambda
z_{1},\lambda z).  \label{Zi2}
\end{equation}%
Here and in what follows we use the notation
\begin{equation}
Z_{\mu }(x,y)=J_{\mu }(y)-\frac{J_{\nu }(x)}{Y_{\nu }(x)}Y_{\mu }(y).
\label{Znu}
\end{equation}

In order to write the solution of the Dirac equation in a more compact form
we introduce the notations%
\begin{equation}
C_{+}^{(J)}=\left(
\begin{array}{c}
C_{+\uparrow }^{(J)} \\
C_{+\downarrow }^{(J)}%
\end{array}%
\right)  \label{C+}
\end{equation}%
and%
\begin{equation}
\widehat{Z}_{\pm }(x,y)=\left(
\begin{array}{cc}
Z_{ma\pm 1/2}(x,y) & 0 \\
0 & Z_{ma\mp 1/2}(x,y)%
\end{array}%
\right) .  \label{Zipm}
\end{equation}%
With these notations, the solution (\ref{psi2}) obeying the boundary
condition at $z=z_{1}$ can be expressed in the form%
\begin{equation}
\psi =z^{(D+1)/2}e^{i\mathbf{kx}-i\omega t}\left(
\begin{array}{c}
\widehat{Z}_{-}(\lambda z_{1},\lambda z)C_{+}^{(J)} \\
\omega ^{-1}\widehat{Z}_{+}(\lambda z_{1},\lambda z)\left( i\lambda
+k_{l}\sigma _{l}\right) C_{+}^{(J)}%
\end{array}%
\right) .  \label{psi3}
\end{equation}%
This solution corresponds to a state of the fermionic field with a given
value of the momentum parallel to the boundary and with a given value of $%
\lambda $. In order to completely specify the solutions we still need an
additional quantum number. This corresponds to fixing the spinor $%
C_{+}^{(J)} $. Here we take $C_{+}^{(J)}=C_{\beta }^{(+)}w^{(\sigma )}$,
where $C_{\beta }^{(+)}$ is a constant and $w^{(\sigma )}$, $\sigma =$ $%
1,\ldots ,N/2$, are one-column matrices of $N/2$ rows, with elements $%
w_{l}^{(\sigma )}=\delta _{l\sigma }$. It can be seen that with this choice,
the solutions (\ref{psi3}), in combination with the corresponding
negative-energy solutions (see below), do form a complete set specified by
the quantum numbers $\beta =(\mathbf{k},\lambda ,\sigma )$. Hence, the
positive-energy mode functions obeying the boundary condition at $z=z_{1}$
have the form%
\begin{equation}
\psi _{\beta }^{(+)}=C_{\beta }^{(+)}z^{\frac{D+1}{2}}e^{i\mathbf{kx}%
-i\omega t}\left(
\begin{array}{c}
\widehat{Z}_{-}(\lambda z_{1},\lambda z)w^{(\sigma )} \\
\frac{1}{\omega }\widehat{Z}_{+}(\lambda z_{1},\lambda z)\left( i\lambda
+k_{l}\sigma _{l}\right) w^{(\sigma )}%
\end{array}%
\right) .  \label{psi+}
\end{equation}

Now we impose the boundary condition on the right boundary, located at $%
z=z_{2}$. From this condition it follows that the eigenvalues of the quantum
number $\lambda $ are roots of the equation%
\begin{equation}
g_{\nu ,\nu -1}(\lambda z_{1},\lambda z_{2})=0,  \label{lambModes}
\end{equation}%
where we have defined the function
\begin{equation}
g_{\nu ,\mu }(x,y)=J_{\nu }(x)Y_{\mu }(y)-J_{\mu }(y)Y_{\nu }(x).
\label{gnumu}
\end{equation}%
Eq.~(\ref{lambModes}) has an infinite number of positive roots. We will
denote them by $\lambda =\lambda _{n}=\gamma _{\nu ,n}/z_{1}$, $n=1,2,\ldots
$.

The coefficient $C_{\beta }^{(+)}$ in (\ref{psi+}) is determined from the
normalization condition%
\begin{equation}
\int d^{D-1}x\int_{z_{1}}^{z_{2}}dz\,\sqrt{|\gamma |}\psi _{\beta
}^{(+)+}\psi _{\beta ^{\prime }}^{(+)}=\delta (\mathbf{k-k}^{\prime })\delta
_{\sigma \sigma ^{\prime }}\delta _{nn^{\prime }},  \label{NormCond}
\end{equation}%
where $\gamma $ is the determinant of the spatial metric, $|\gamma
|=(a/z)^{D}$. By using a standard result for the integral involving the
square of the cylinder functions (see, for instance, \cite{Prud86}), for the
normalization coefficient we find%
\begin{equation}
|C_{\beta }^{(+)}|^{2}=\frac{\pi ^{2}\lambda Y_{\nu }^{2}(\lambda z_{1})}{%
4(2\pi )^{D-1}a^{D}z_{1}}T_{\nu }(\eta ,\lambda z_{1}),  \label{Cbet}
\end{equation}%
where we have introduced the notations%
\begin{equation}
\eta =z_{2}/z_{1},  \label{eta}
\end{equation}
and%
\begin{equation}
T_{\nu }(\eta ,x)=x\left[ \frac{J_{\nu }^{2}(x)}{J_{\nu -1}^{2}(\eta x)}-1%
\right] ^{-1}.  \label{Tnu}
\end{equation}%
This finishes the construction of the positive-energy mode functions.

The negative-energy mode functions can be obtained in a similar way. They
have the form:%
\begin{equation}
\psi _{\beta }^{(-)}=C_{\beta }^{(-)}z^{\frac{D+1}{2}}e^{i\mathbf{kx}%
-i\omega t}\left(
\begin{array}{c}
\frac{1}{\omega }\widehat{Z}_{-}(\lambda z_{1},\lambda z)\left( i\lambda
-k_{l}\sigma _{l}\right) w^{(\sigma )} \\
\widehat{Z}_{+}(\lambda z_{1},\lambda z)w^{(\sigma )}%
\end{array}%
\right) ,  \label{psi-}
\end{equation}%
where $|C_{\beta }^{(-)}|^{2}$ is given by the same expression (\ref{Cbet}).
As in the case of the positive-energy modes, for the eigenvalues of $\lambda
$ one has $\lambda _{n}=\gamma _{\nu ,n}/z_{1}$.

\section{Fermionic condensate}

\label{sec:FC}

In this section we consider the FC defined as the VEV $\left\langle
0\right\vert \bar{\psi}\psi \left\vert 0\right\rangle \equiv \langle \bar{%
\psi}\psi \rangle $, where $\left\vert 0\right\rangle $ corresponds to the
vacuum state and $\bar{\psi}=\psi ^{+}\gamma ^{(0)}$ is the Dirac adjoint.
Note that the Dirac adjoint is defined through the flat spacetime matrix $%
\gamma ^{(0)}$. In addition to describing the physical structure of the
quantum field at a given point, the FC plays an important role in models of
dynamical chiral symmetry breaking (see the reviews \cite{Inag97}, for
chiral symmetry breaking on curved spacetime with nontrivial topology, and
\cite{Flac11} for recent developments).

Expanding the field operator in terms of the complete set of positive- and
negative-energy mode functions $\{\psi _{\beta }^{(+)},\psi _{\beta
}^{(-)}\} $, and using the anticommutation relations for the annihilation
and creation operators, we find the mode-sum formula for the fermionic
condensate:
\begin{equation}
\langle \bar{\psi}\psi \rangle =\sum_{\beta }\bar{\psi}_{\beta }^{(-)}\psi
_{\beta }^{(-)},  \label{FC}
\end{equation}%
where
\begin{equation}
\sum_{\beta }=\int d\mathbf{k}\sum_{\sigma =1}^{N/2}\sum_{n=1}^{\infty }.
\label{Sumbet}
\end{equation}%
The expression on the right-hand side (rhs) of (\ref{FC}) is divergent and,
to make sense of it, some regularization procedure is needed. Here we assume
that a cutoff function is present, without explicitly writing it. The
special form of this function will not be important in the further
discussion.

Substituting the mode functions (\ref{psi-}) into (\ref{FC}) after some
transformations the fermionic condensate can be expressed as%
\begin{equation}
\langle \bar{\psi}\psi \rangle =-\frac{Na^{-D}z^{D+1}}{16(2\pi
)^{D-3}z_{1}^{2}}\int d\mathbf{k}\sum_{n=1}^{\infty }\frac{\gamma _{\nu
,n}^{2}T_{\nu }(\eta ,\gamma _{\nu ,n})}{\sqrt{\gamma _{\nu
,n}^{2}+k^{2}z_{1}^{2}}}g_{\nu ,\nu }(\gamma _{\nu ,n},\gamma _{\nu
,n}z/z_{1})g_{\nu ,\nu -1}(\gamma _{\nu ,n},\gamma _{\nu ,n}z/z_{1}),
\label{FC1}
\end{equation}%
the function $g_{\nu ,\mu }(x,y)$ being defined by (\ref{gnumu}). As the
roots $\gamma _{\nu ,n}$ are given implicitly, the form (\ref{FC1}) for the
FC is not convenient for the investigation of the effects induced by the
boundaries. In addition, the terms in the series are highly oscillatory for
large values of $n$.

A more convenient form of the mode sum for the FC is obtained by using the
summation formula%
\begin{equation}
\sum_{n=1}^{\infty }T_{\nu }(\eta ,\gamma _{\nu ,n})h(\gamma _{\nu ,n})=%
\frac{2}{\pi ^{2}}\int_{0}^{\infty }\frac{h(x)dx}{J_{\nu }^{2}(x)+Y_{\nu
}^{2}(x)}+\frac{1}{2\pi }\int_{0}^{\infty }dx\,\Omega _{\nu }^{(1)}(x,\eta x)%
\left[ h(ix)+h(-ix)\right] ,  \label{SumForm}
\end{equation}%
where%
\begin{equation}
\Omega _{\nu }^{(1)}(x,y)=\frac{K_{\nu -1}(y)/K_{\nu }(x)}{K_{\nu }(x)I_{\nu
-1}(y)+I_{\nu }(x)K_{\nu -1}(y)},  \label{Om1}
\end{equation}%
and $I_{\nu }(x)$, $K_{\nu }(x)$ are modified Bessel functions. This formula
is derived in Ref. \cite{Saha01} by using the generalized Abel-Plana formula
(see also \cite{SahaBook}). The corresponding conditions on the function $%
h(u)$, analytic on the right half plane of the complex variable $u$, can be
found in \cite{Saha01}. Applying expression (\ref{SumForm}) to the series
over $n$ in (\ref{FC1}), after the integration over the angular part of $%
\mathbf{k}$, and introducing a new integration variable $x=kz_{1}$, the FC
can be written as%
\begin{eqnarray}
\langle \bar{\psi}\psi \rangle &=&\langle \bar{\psi}\psi \rangle _{1}+\frac{%
8N(z/z_{1})^{D+1}a^{-D}}{(4\pi )^{(D+1)/2}\Gamma ((D-1)/2)}\int_{0}^{\infty
}dx\,x^{D-2}\mathbf{\,}  \notag \\
&&\times \int_{x}^{\infty }du\,\frac{u^{2}\Omega _{\nu }^{(1)}(u,\eta u)}{%
\sqrt{u^{2}-x^{2}}}G_{\nu ,\nu }(u,uz/z_{1})G_{\nu ,\nu -1}(u,uz/z_{1}).
\label{FC3}
\end{eqnarray}%
where we have introduced the notation%
\begin{equation}
G_{\nu ,\mu }(x,y)=I_{\nu }(x)K_{\mu }(y)-(-1)^{\nu -\mu }I_{\mu }(y)K_{\nu
}(x).  \label{Gnumu}
\end{equation}%
The first term on the rhs of (\ref{FC3}) comes from the first integral in (%
\ref{SumForm}) and it is given by%
\begin{eqnarray}
\langle \bar{\psi}\psi \rangle _{1} &=&-\frac{N(z/z_{1})^{D+1}a^{-D}}{(4\pi
)^{(D-1)/2}\Gamma ((D-1)/2)}\int_{0}^{\infty }dx\,x^{D-2}\int_{0}^{\infty }du
\notag \\
&&\times \frac{u^{2}}{\sqrt{u^{2}+x^{2}}}\frac{g_{\nu ,\nu
}(u,uz/z_{1})g_{\nu ,\nu -1}(u,uz/z_{1})}{J_{\nu }^{2}(u)+Y_{\nu }^{2}(u)}.
\label{FCS1}
\end{eqnarray}%
We now first consider this term.

\subsection{Condensate for the geometry of a single boundary}

The second term on the rhs of (\ref{FC3}) is finite for $z<z_{2}$, in the
absence of the cutoff function. This term vanishes in the limit $%
z_{2}\rightarrow \infty $, whereas $\langle \bar{\psi}\psi \rangle _{1}$
does not depend on $z_{2}$. This allows us to interpret the part (\ref{FCS1}%
) as the FC in the region $z>z_{1}$ for the geometry of a single boundary at
$z=z_{1}$ when the other boundary is absent. This can also be seen by direct
evaluation of the FC using Eq.~(\ref{FC}). The corresponding positive-energy
mode functions are given by (\ref{psi+}), where now the spectrum for $%
\lambda $ is continuous. Consequently, in (\ref{FC}) we have $\sum_{\beta
}=\int d\mathbf{k}\int_{0}^{\infty }d\lambda \sum_{\sigma =1}^{N/2}$. The
normalization coefficient is determined by the condition which is obtained
from (\ref{NormCond}) by the replacements $\int_{z_{1}}^{z_{2}}dz\rightarrow
\int_{z_{1}}^{\infty }dz$ and $\delta _{nn^{\prime }}\rightarrow \delta
(\lambda -\lambda ^{\prime })$. In this way, we can see that%
\begin{equation}
|C_{\beta }^{(+)}|^{2}=\frac{a^{-D}\lambda }{2(2\pi )^{D-1}}\left[ 1+\frac{%
J_{\nu }^{2}(\lambda z_{1})}{Y_{\nu }^{2}(\lambda z_{1})}\right] ^{-1}.
\label{C+1b}
\end{equation}%
Similarly, the negative-energy modes have the form (\ref{psi-}) with $%
|C_{\beta }^{(-)}|^{2}$ given by the same expression (\ref{C+1b}).
Substituting these mode functions into the formula (\ref{FC}), the
expression (\ref{FCS1}) is obtained for the single boundary part.

For further transformation of such expression we use the identity%
\begin{equation}
\frac{g_{\nu ,\nu }(u,y)g_{\nu ,\nu -1}(u,y)}{J_{\nu }^{2}(u)+Y_{\nu }^{2}(u)%
}=J_{\nu }(y)J_{\nu -1}(y)-\frac{1}{2}\sum_{s=1,2}\frac{J_{\nu }(u)}{H_{\nu
}^{(s)}(u)}H_{\nu }^{(s)}(y)H_{\nu -1}^{(s)}(y),  \label{Ident1}
\end{equation}%
the $H_{\nu }^{(s)}(y)$, $s=1,2$, being Hankel functions. Substituting (\ref%
{Ident1}) into (\ref{FCS1}), in the integral over $u$ with the second term
on the rhs of (\ref{Ident1}) we rotate the integration contour in the
complex plane $u$ by the angle $\pi /2$ ($-\pi /2$) for the term with $s=1$ (%
$s=2$). Introducing the modified Bessel functions, the FC is then decomposed
into%
\begin{equation}
\langle \bar{\psi}\psi \rangle _{1}=\langle \bar{\psi}\psi \rangle
_{0}+\langle \bar{\psi}\psi \rangle _{1}^{\mathrm{(b)}},  \label{FCSdec}
\end{equation}%
with%
\begin{equation}
\langle \bar{\psi}\psi \rangle _{0}=-\frac{(4\pi )^{(1-D)/2}N}{\Gamma
((D-1)/2)a^{D}}\int_{0}^{\infty }dx\,x^{D-2}\int_{0}^{\infty }dy\frac{%
y^{2}J_{\nu }(y)}{\sqrt{y^{2}+x^{2}}}J_{\nu -1}(y)  \label{FC0}
\end{equation}%
and%
\begin{eqnarray}
\langle \bar{\psi}\psi \rangle _{1}^{\mathrm{(b)}} &=&-\frac{%
N(z/z_{1})^{D+1}a^{-D}}{(4\pi )^{(D-1)/2}\Gamma ((D-1)/2)}\int_{0}^{\infty
}dx\,x^{D-2}\mathbf{\,}  \notag \\
&&\times \frac{2}{\pi }\int_{x}^{\infty }du\frac{u^{2}}{\sqrt{u^{2}-x^{2}}}%
\frac{I_{\nu }(u)}{K_{\nu }(u)}K_{\nu }(uz/z_{1})K_{\nu -1}(uz/z_{1}).
\label{FCS2}
\end{eqnarray}%
The second part (\ref{FCS2}) is finite for $z>z_{1}$ and the cutoff function
need be kept in the first part (\ref{FC0}) only. Note that the latter does
not depend on $z$. The term $\langle \bar{\psi}\psi \rangle _{1}^{\mathrm{(b)%
}}$ vanishes in the limit $z\rightarrow \infty $ and, hence, the term $%
\langle \bar{\psi}\psi \rangle _{0}$ can be interpreted as the pure AdS part
of the FC when the boundaries are absent. The property that this term is
uniform is just a consequence of the maximal symmetry both of AdS spacetime
and of the vacuum state we have chosen. With the representation (\ref{FCSdec}%
), the renormalization of the FC outside the boundary is thus reduced to the
one corresponding to AdS spacetime when the boundaries are absent.

The expression (\ref{FCS2}) for the boundary-induced part can be further
simplified by using the integration formula%
\begin{equation}
\int_{0}^{\infty }dx\,x^{D-2}\int_{x}^{\infty }du\frac{f(u)}{\sqrt{%
u^{2}-x^{2}}}=\sqrt{\pi }\frac{\Gamma ((D-1)/2)}{2\Gamma (D/2)}%
\int_{0}^{\infty }dr\,r^{D-2}\,f(r).  \label{IntForm1}
\end{equation}%
This is obtained by introducing a new integration variable $y=\sqrt{%
u^{2}-x^{2}}$, passing to polar coordinates on the plane $(x,y)$ and then
integrating over the polar angle. Now, the expression for the FC induced in
the region $z>z_{j}$ by a single boundary, located at $z=z_{j}$, simply reads%
\begin{equation}
\langle \bar{\psi}\psi \rangle _{j}^{\mathrm{(b)}}=-\frac{2N(z/z_{j})^{D+1}}{%
A_{D}a^{D}}\int_{0}^{\infty }dx\,x^{D}\frac{I_{\nu }(x)}{K_{\nu }(x)}K_{\nu
}(xz/z_{j})K_{\nu -1}(xz/z_{j}),  \label{FCb}
\end{equation}%
where we have introduced the notation%
\begin{equation}
A_{D}=(4\pi )^{D/2}\Gamma (D/2).  \label{AD}
\end{equation}%
Note that the boundary-induced part (\ref{FCb}) is a function of the ratio $%
z/z_{j}$ alone. By taking into account that $z/z_{j}=e^{(y-y_{j})/a}$, we
see that for a given distance from the boundary, $y-y_{j}$, the quantity $%
\langle \bar{\psi}\psi \rangle _{1}^{\mathrm{(b)}}$ does not depend on the
location of the boundary. Again, this is a consequence of the maximal
symmetry of the AdS bulk. It follows from (\ref{FCb}) that $\langle \bar{\psi%
}\psi \rangle _{1}^{\mathrm{(b)}}$ is negative.

The expression (\ref{FCb}) gives the boundary-induced part in the region $%
z>z_{j}$. In order to find the condensate induced by a single boundary in
the region to the left of the boundary, we take the limit $z_{1}\rightarrow
0 $ in the general expression (\ref{FC3}) for the geometry with two
boundaries. Introducing in (\ref{FCb}), with $j=1$, a new integration
variable, $y=x/z_{1}$, we see that in this limit $\langle \bar{\psi}\psi
\rangle _{1}^{\mathrm{(b)}}$ vanishes and $\langle \bar{\psi}\psi \rangle
_{1}$ reduces to $\langle \bar{\psi}\psi \rangle _{0}$. In the second term
on the rhs of (\ref{FC3}), passing to the new integration variables $%
u^{\prime }=u/z_{1}$ and $x^{\prime }=x/z_{1}$, the limit $z_{1}\rightarrow
0 $ is readily evaluated. As a result, from (\ref{FC3}) we obtain the FC in
the region $z<z_{2}$ for the geometry of a single boundary at $z=z_{2}$. For
the geometry of a single boundary at $z=z_{j}$, the FC in the region $%
z<z_{j} $ is expressed as $\langle \bar{\psi}\psi \rangle _{j}=\langle \bar{%
\psi}\psi \rangle _{0}+\langle \bar{\psi}\psi \rangle _{j}^{\mathrm{(b)}}$,
with the boundary-induced part being%
\begin{equation}
\langle \bar{\psi}\psi \rangle _{j}^{\mathrm{(b)}}=-\frac{2N(z/z_{j})^{D+1}}{%
A_{D}a^{D}}\int_{0}^{\infty }dx\,x^{D}\,\frac{K_{\nu -1}(x)}{I_{\nu -1}(x)}%
I_{\nu }(xz/z_{j})I_{\nu -1}(xz/z_{j}).  \label{FCbLeft}
\end{equation}%
This expression is finite for points away from the boundary and vanishes at
the AdS boundary. Similar to the case of the region to the right, the FC
defined by (\ref{FCbLeft}) does not depend on the location of the boundary
for a fixed distance from it. Note that the FC is not symmetric with respect
to the boundary, the reason for this being that, though the background
spacetime is homogeneous, the boundary $z=z_{j}$ has a nonzero extrinsic
curvature tensor and the two sides of the boundary are not equivalent. Here
the situation is similar to that for curved boundaries in the Minkowski bulk.

Let us consider the asymptotic behavior of the single boundary-induced part.
At large distances from the boundary, $z\gg z_{j}$, we introduce in (\ref%
{FCb}) a new integration variable $y=xz/z_{1}$ and expand the integrand by
using the formulas for the modified Bessel functions for small values of the
argument. To leading order, the remaining integral which involves the
product of two MacDonald functions is evaluated by using a formula from \cite%
{Prud86}. In this way, one finds that%
\begin{equation}
\langle \bar{\psi}\psi \rangle _{j}^{\mathrm{(b)}}\approx -\frac{%
Na^{-D}(z_{j}/z)^{2\nu }}{2^{D+2\nu +1}\pi ^{(D-1)/2}}\frac{D\Gamma
(D/2+2\nu )\Gamma (D/2+\nu )}{\nu \Gamma (D/2+\nu +1/2)\Gamma ^{2}(\nu )}.
\label{FCfar}
\end{equation}%
For the region to the left of the boundary, assuming $z\ll z_{j}$, by direct
expansion of the integrand in (\ref{FCbLeft}), to leading order one gets%
\begin{equation}
\langle \bar{\psi}\psi \rangle _{j}^{\mathrm{(b)}}\approx -\frac{2^{2-2\nu
}N(z/z_{j})^{D+2\nu }}{A_{D}a^{D}\nu \Gamma ^{2}(\nu )}\int_{0}^{\infty
}dx\,x^{D+2\nu -1}\,\frac{K_{\nu -1}(x)}{I_{\nu -1}(x)}.  \label{FCfar2}
\end{equation}%
This expression gives the asymptotic behavior of the FC near the AdS
boundary. Note that both limits, (\ref{FCfar}) and (\ref{FCfar2}),
correspond to distances from the boundary much larger than the curvature
scale for the background spacetime: $|y-y_{j}|\gg a$. As we see, in these
regions the boundary-induced part decays exponentially with the distance
from the boundary.

For points near the boundary, $|1-z/z_{j}|\ll 1$, the dominant contributions
to the integrals in (\ref{FCb}) and (\ref{FCbLeft}) come from large values
of $x$. By using the asymptotic expressions for the modified Bessel
functions for large values of the argument, to leading order we find%
\begin{equation}
\langle \bar{\psi}\psi \rangle _{j}^{\mathrm{(b)}}\approx -\frac{N\Gamma
((D+1)/2)}{(4\pi )^{(D+1)/2}|y-y_{j}|^{D}}.  \label{FCnear}
\end{equation}%
This leading term coincides with the corresponding one for the boundary in
Minkowski spacetime.

\subsection{FC in the region between the two boundaries}

Now we return to the geometry with two boundaries. Using of the integration
formulas (\ref{IntForm1}) in (\ref{FC3}), the FC in the region $%
z_{1}<z<z_{2} $ can be expressed in the form%
\begin{eqnarray}
\langle \bar{\psi}\psi \rangle &=&\langle \bar{\psi}\psi \rangle
_{0}+\langle \bar{\psi}\psi \rangle _{1}^{\mathrm{(b)}}+\frac{%
2N(z/z_{1})^{D+1}}{A_{D}a^{D}}\int_{0}^{\infty }dx\,x^{D}\,  \notag \\
&&\times \Omega _{\nu }^{(1)}(x,\eta x)G_{\nu ,\nu }(x,xz/z_{1})G_{\nu ,\nu
-1}(x,xz/z_{1}),  \label{FC4}
\end{eqnarray}%
where $\langle \bar{\psi}\psi \rangle _{1}^{\mathrm{(b)}}$ is given by (\ref%
{FCb}) with $j=1$. For points outside the boundaries renormalization is only
needed for the first term on the rhs. In formula (\ref{FC4}), the functions $%
\Omega _{\nu }^{(1)}(x,y)$ and $G_{\nu ,\nu -1}(x,y)$ are positive. Using
the result that the function $I_{\nu }(x)/K_{\nu }(x)$ is a monotonically
increasing in terms of $x>0$, we see that the function $G_{\nu ,\nu }(x,y)$
is negative for $x<y$ and positive for $x>y$. Hence, the last term in (\ref%
{FC4}) is negative. Combining this with the result $\langle \bar{\psi}\psi
\rangle _{1}^{\mathrm{(b)}}<0$, we conclude that the boundary-induced part
of the FC in the region between the boundaries is negative.

It can be checked that, in the region between the boundaries, the FC can
also be written as%
\begin{eqnarray}
\langle \bar{\psi}\psi \rangle &=&\langle \bar{\psi}\psi \rangle
_{0}+\langle \bar{\psi}\psi \rangle _{2}^{\mathrm{(b)}}-\frac{%
2N(z/z_{2})^{D+1}}{A_{D}a^{D}}\int_{0}^{\infty }dx\,x^{D}\,  \notag \\
&&\times \Omega _{\nu }^{(2)}(x/\eta ,x)G_{\nu -1,\nu -1}(x,xz/z_{2})G_{\nu
-1,\nu }(x,xz/z_{2}),  \label{FC5}
\end{eqnarray}%
with%
\begin{equation}
\Omega _{\nu }^{(2)}(x,y)=\frac{I_{\nu }(x)/I_{\nu -1}(y)}{K_{\nu }(x)I_{\nu
-1}(y)+I_{\nu }(x)K_{\nu -1}(y)}.  \label{Om2}
\end{equation}%
In (\ref{FC5}), $\langle \bar{\psi}\psi \rangle _{2}^{\mathrm{(b)}}$ is
given by (\ref{FCbLeft}) with $j=2$ and the last term is induced by the
boundary at $z=z_{1}$. The latter vanishes in the limit $z_{1}\rightarrow 0$%
, it is finite for $z=z_{2}$, and diverges on the left boundary. Divergences
occurring in this term are the same as those for a single boundary at $%
z=z_{2}$. All functions in the integrand of the last term in (\ref{FC5}) are
positive. From the discussion above it follows that, for the region between
the boundaries, if we express the FC in the form%
\begin{equation}
\langle \bar{\psi}\psi \rangle =\langle \bar{\psi}\psi \rangle
_{0}+\sum_{j=1,2}\langle \bar{\psi}\psi \rangle _{j}^{\mathrm{(b)}}+\Delta
\langle \bar{\psi}\psi \rangle ,  \label{FCInt}
\end{equation}%
then the last (interference) term is finite everywhere, including the points
on the boundaries. The surface divergences are contained in the single
boundary parts. At large distances between the boundaries, as compared with
the curvature radius of the background spacetime, $\left( y_{2}-y_{1}\right)
\gg a$, the interference part is exponentially suppressed.

Consider now the Minkowskian limit of the expression for the FC in the
region between the boundaries. This corresponds to $ma\gg 1$ for a fixed
value of $y$, hence, one has $\nu \gg 1$ and $z/z_{j}\approx 1+(y-y_{j})/a$.
Introducing in the formulae above a new integration variable $u=x/\nu $ and
using the uniform asymptotic expansions for the modified Bessel functions
for large values of the order, after some transformations, to leading order
we get
\begin{eqnarray}
\langle \bar{\psi}\psi \rangle &\approx &\langle \bar{\psi}\psi \rangle _{%
\mathrm{(M)}}=-\frac{N}{A_{D}}\int_{m}^{\infty }dx\,\frac{%
(x^{2}-m^{2})^{D/2-1}}{\frac{x+m}{x-m}e^{2x(y_{2}-y_{1})}+1}  \notag \\
&&\times \{(m+x)[e^{2x(y-y_{1})}+e^{2x(y_{2}-y)}]-2m\},  \label{FCM}
\end{eqnarray}%
where $\langle \bar{\psi}\psi \rangle _{\mathrm{(M)}}$ is the FC for
boundaries in a Minkowski bulk. The expression (\ref{FCM}) for $\langle \bar{%
\psi}\psi \rangle _{\mathrm{(M)}}$ is a special case of a more general
formula derived in \cite{Eliz11} for the Minkowski bulk with compact spatial
dimensions. The fermionic condensate for a massless field has been
considered in \cite{Lutk84}.

\section{VEV of the energy-momentum tensor}

\label{sec:EMT}

The VEV of the energy-momentum tensor is another important local
characteristic of the fermionic vacuum. In order to find this VEV we use the
mode-sum formula
\begin{equation}
\langle T_{\mu \alpha }\rangle =\frac{i}{2}\int d\mathbf{k}\sum_{\sigma
=1}^{N/2}\sum_{n=1}^{\infty }[\bar{\psi}_{\beta }^{(-)}\gamma _{(\mu }\nabla
_{\alpha )}\psi _{\beta }^{(-)}-(\nabla _{(\mu }\bar{\psi}_{\beta
}^{(-)})\gamma _{\alpha )}\psi _{\beta }^{(-)}]\ ,  \label{modesum}
\end{equation}%
where the brackets denote symmetrization over the indices enclosed. Note
that for the covariant derivative of the Dirac adjoint field one has $\nabla
_{\mu }\bar{\psi}_{\beta }^{(-)}=\partial _{\mu }\bar{\psi}_{\beta }^{(-)}-%
\bar{\psi}_{\beta }^{(-)}\Gamma _{\mu }$. We see that the spin connection
appears in the expression for the VEV in the form $\gamma _{(\mu }\Gamma
_{\nu )}+\Gamma _{(\mu }\gamma _{\nu )}$. By using the expressions (\ref%
{SpinCon}) it can be seen that this combination vanishes. Hence, in the
evaluation of the VEVs we can make the replacement $\nabla _{\nu
}\rightarrow \partial _{\nu }$.

Substituting (\ref{psi-}) for the negative-energy mode functions into (\ref%
{modesum}), it can be seen that the off-diagonal components vanish. For the
VEVs of the diagonal components we find the following expressions (no
summation over $\mu $)%
\begin{equation}
\langle T_{\mu }^{\mu }\rangle =\frac{N\pi ^{2}a^{-D-1}(z/z_{1})^{D+2}}{%
4(4\pi )^{(D-1)/2}\Gamma ((D-1)/2)}\int_{0}^{\infty }du\,u^{D-2}\mathbf{\,}%
\sum_{n=1}^{\infty }\frac{\gamma _{\nu ,n}T_{\nu }(\eta ,\gamma _{\nu ,n})}{%
\sqrt{\gamma _{\nu ,n}^{2}+u^{2}}}f_{\nu }^{(\mu )}(\gamma _{\nu ,n},\gamma
_{\nu ,n}z/z_{1}),  \label{EMT1}
\end{equation}%
with the notations%
\begin{eqnarray}
f_{\nu }^{(0)}(x,y) &=&-\left( x^{2}+u^{2}\right) \left[ g_{\nu ,\nu
}^{2}(x,y)+g_{\nu ,\nu -1}^{2}(x,y)\right] ,  \notag \\
f_{\nu }^{(l)}(x,y) &=&\frac{u^{2}}{D-1}\left[ g_{\nu ,\nu }^{2}(x,y)+g_{\nu
,\nu -1}^{2}(x,y)\right] ,  \label{fl} \\
f_{\nu }^{(D)}(x,y) &=&x^{2}\left[ g_{\nu ,\nu }^{2}(x,y)+g_{\nu ,\nu
-1}^{2}(x,y)-\frac{2\nu -1}{y}g_{\nu ,\nu }(x,y)g_{\nu ,\nu -1}(x,y)\right] ,
\notag
\end{eqnarray}%
and $l=1,\ldots ,D-1$. As in the case of the FC, here we assume the presence
of a cutoff function which makes the expressions (\ref{EMT1}) finite.

Further transformations of the VEVs proceeds similarly to those in the case
of the FC. First, the series over $n$ in (\ref{EMT1}) is transformed via the
summation formula (\ref{SumForm}). Next, for the part corresponding to the
second term on the rhs of (\ref{SumForm}) we use the integration formula (%
\ref{IntForm1}). In this way, the VEVs are expressed as (no summation over $%
\mu $)
\begin{equation}
\langle T_{\mu }^{\mu }\rangle =\langle T_{\mu }^{\mu }\rangle _{1}+\frac{%
N(z/z_{1})^{D+2}}{A_{D}a^{D+1}}\int_{0}^{\infty }dx\,x^{D+1}\Omega _{\nu
}^{(1)}(x,\eta x)F_{1\nu }^{(\mu )}(x,xz/z_{1}),  \label{EMT2}
\end{equation}%
where the notations
\begin{eqnarray}
F_{1\nu }^{(l)}(x,y) &=&\frac{1}{D}\left[ G_{\nu ,\nu }^{2}(x,y)-G_{\nu ,\nu
-1}^{2}(x,y)\right] ,  \notag \\
F_{1\nu }^{(D)}(x,y) &=&G_{\nu ,\nu -1}^{2}(x,y)-G_{\nu ,\nu }^{2}(x,y)+%
\frac{2\nu -1}{y}G_{\nu ,\nu }(x,y)G_{\nu ,\nu -1}(x,y),  \label{FD}
\end{eqnarray}%
are introduced with $l=0,\ldots ,D-1$. In (\ref{EMT2}), the first term comes
from the first integral in the summation formula (\ref{SumForm}) and it has
the form (no summation over $\mu $)%
\begin{eqnarray}
\langle T_{\mu }^{\mu }\rangle _{1} &=&\frac{Na^{-D-1}(z/z_{1})^{D+2}}{%
2(4\pi )^{(D-1)/2}\Gamma ((D-1)/2)}\int_{0}^{\infty }du\,u^{D-2}\mathbf{\,}
\notag \\
&&\times \int_{0}^{\infty }dx\frac{x}{\sqrt{x^{2}+u^{2}}}\frac{f_{\nu
}^{(\mu )}(x,xz/z_{1})}{J_{\nu }^{2}(x)+Y_{\nu }^{2}(x)}.  \label{EMTs1}
\end{eqnarray}%
This term corresponds to the VEV of the energy-momentum tensor in the region
$z>z_{1}$ for the geometry of a single boundary at $z=z_{1}$. The second
term on the rhs of (\ref{EMT2}) is induced by the presence of the second
boundary at $z=z_{2}$. The latter is finite for $z_{1}\leq z<z_{2} $ and
renormalization is needed for the first term, only.

The single boundary part (\ref{EMTs1}) could also be directly obtained by
using the the corresponding mode functions. The latter are given by (\ref%
{psi+}) and (\ref{psi-}), with $0\leq \lambda <0$ and with the normalization
coefficient (\ref{C+1b}).

The transformation of the part $\langle T_{\mu }^{\mu }\rangle _{1}$ is also
similar to that for corresponding term in the FC. Using (\ref{Ident1}) and
the similar identity
\begin{equation}
\frac{g_{\nu ,\mu }^{2}(x,y)}{J_{\nu }^{2}(x)+Y_{\nu }^{2}(x)}=J_{\mu
}^{2}(y)-\frac{1}{2}\sum_{s=1,2}\frac{J_{\nu }(x)}{H_{\nu }^{(s)}(x)}H_{\mu
}^{(s)2}(y),  \label{Ident2}
\end{equation}%
the integrand in (\ref{EMTs1}) can be decomposed into parts containing
Bessel and Hankel functions, with the argument $xz/z_{1}$. Then, rotating
the integration contour over $x$ by an angle $\pi /2$ ($-\pi /2$), for the
part with the function $H_{\mu }^{(1)}(xz/z_{1})$ ($H_{\mu }^{(2)}(xz/z_{1})$%
), we get (no summation over $\mu $)%
\begin{equation}
\langle T_{\mu }^{\mu }\rangle _{1}=\langle T_{\mu }^{\mu }\rangle
_{0}+\langle T_{\mu }^{\mu }\rangle _{1}^{\mathrm{(b)}},  \label{EMTs2}
\end{equation}%
where the two terms are respectively given by%
\begin{equation}
\langle T_{\mu }^{\mu }\rangle _{0}=\frac{(4\pi )^{-(D-1)/2}N}{2\Gamma
((D-1)/2)a^{D+1}}\int_{0}^{\infty }dx\,x^{D-2}\int_{0}^{\infty }du\frac{%
uf_{0\nu }^{(\mu )}(u)}{\sqrt{u^{2}+x^{2}}},  \label{Tmu0}
\end{equation}%
and
\begin{equation}
\langle T_{\mu }^{\mu }\rangle _{1}^{\mathrm{(b)}}=-\frac{N(z/z_{1})^{D+2}}{%
A_{D}a^{D+1}}\int_{0}^{\infty }dx\,x^{D+1}\frac{I_{\nu }(x)}{K_{\nu }(x)}%
S_{1\nu }^{(\mu )}(xz/z_{1}).  \label{EMTb1}
\end{equation}%
In (\ref{Tmu0}), the expressions for $f_{0\nu }^{(\mu )}(y)$ are obtained
from the corresponding expressions for $f_{\nu }^{(\mu )}(x,y)$ by the
replacement $g_{\nu ,\alpha }(x,y)\rightarrow J_{\alpha }(y)$, and in (\ref%
{EMTb1}) we have defined%
\begin{eqnarray}
S_{1\nu }^{(l)}(x) &=&\frac{1}{D}\left[ K_{\nu }^{2}(x)-K_{\nu -1}^{2}(x)%
\right] ,  \notag \\
S_{1\nu }^{(D)}(x) &=&K_{\nu -1}^{2}(x)-K_{\nu }^{2}(x)+\frac{2\nu -1}{x}%
K_{\nu }(x)K_{\nu -1}(x),  \label{Snu}
\end{eqnarray}%
with $l=0,\ldots ,D-1$. As we see, in both cases of single and double
boundary geometries the energy density is equal to the stresses along the
directions parallel to the boundaries. This property is related to the
symmetry of the problem under consideration.

$\langle T_{\mu }^{\mu }\rangle _{1}^{\mathrm{(b)}}$ vanishes in the limit $%
z_{1}\rightarrow 0$ (for the corresponding asymptotic behavior see below)
while $\langle T_{\alpha }^{\mu }\rangle _{0}$ can be identified with the
VEV of the energy-momentum tensor in the boundary-free AdS bulk. With the
decomposition (\ref{EMTs2}) and for points away from the boundary,
renormalization is required for this last part only. Because of the maximal
symmetry of the background geometry, the renormalized VEV does not depend on
the spacetime position and is completely determined by the trace: $\langle
T_{\alpha }^{\mu }\rangle _{0}=\langle T_{\sigma }^{\sigma }\rangle
_{0}\delta _{\alpha }^{\mu }/(D+1)$. For the case $D=3$ this VEV was
investigated in \cite{Camp92} using the zeta-function techniques and also
Pauli-Villars regularization. In what follows we specifically discuss the
effects induced by the boundaries.

For a massless field one has $\nu =1/2$ and from (\ref{EMTb1}) it
can be easily seen that $\langle T_{\alpha }^{\mu }\rangle
_{1}^{\mathrm{(b)}}=0$ for $z>z_{1}$. We could obtain this result
directly, by taking into account that for a massless fermionic
field the problem is conformally related to the corresponding
problem for a single boundary in the Minkowski bulk and for the
latter geometry the VEV of the energy-momentum tensor vanishes. In
the region between two boundaries, by making use of the expressions $%
G_{1/2,1/2}(x,y)=(xy)^{-1/2}\sinh (x-y)$ and $G_{1/2,-1/2}(x,y)=(xy)^{-1/2}%
\cosh (x-y)$, from (\ref{EMT2}) we get (no summation over $\mu $)
\begin{equation}
\langle T_{\mu }^{\mu }\rangle =\langle T_{\mu }^{\mu }\rangle
_{0}+\left( \frac{z}{a}\right) ^{D+1}\langle T_{\mu }^{\mu
}\rangle _{\mathrm{(M)},m=0}. \label{EMTm0}
\end{equation}%
where%
\begin{equation}
\langle T_{\mu }^{\mu }\rangle
_{\mathrm{(M)},m=0}=-\frac{N(1-2^{-D})\Gamma ((D+1)/2)}{(4\pi
)^{(D+1)/2}(z_{2}-z_{1})^{D+1}}\zeta _{\mathrm{R}}(D+1),
\label{EMTMm0}
\end{equation}%
for $\mu =0,\ldots ,D-1$, and $\langle T_{D}^{D}\rangle _{\mathrm{(M)}%
,m=0}=-D\langle T_{0}^{0}\rangle _{\mathrm{(M)},m=0}$. In (\ref{EMTm0}), $%
\zeta _{\mathrm{R}}(x)$ is the Riemann zeta function. Here,
$\langle T_{\alpha }^{\mu }\rangle _{\mathrm{(M)},m=0}$ is the
corresponding VEV for two boundaries at $z=z_{1}$ and $z=z_{2}$ in
Minkowski spacetime.   Of course, (\ref{EMTm0}) shows the standard
conformal relation between the problems in AdS and Minkowski
bulks. Note that for a massless field the boundary-free part
$\langle T_{\alpha }^{\mu }\rangle _{0}$ is completely determined
by the trace anomaly.

We can write the VEV of the energy-momentum tensor in an
alternative form (no summation over $\mu $):%
\begin{equation}
\langle T_{\mu }^{\mu }\rangle =\langle T_{\mu }^{\mu }\rangle _{0}+\langle
T_{\mu }^{\mu }\rangle _{2}^{\mathrm{(b)}}+\frac{N(z/z_{2})^{D+2}}{%
A_{D}a^{D+1}}\int_{0}^{\infty }dx\,x^{D+1}\Omega _{\nu }^{(2)}(x/\eta
,x)F_{2\nu }^{(\mu )}(x,xz/z_{2}),  \label{EMT}
\end{equation}%
where the term
\begin{equation}
\langle T_{\mu }^{\mu }\rangle _{2}^{\mathrm{(b)}}=-\frac{N(z/z_{2})^{D+2}}{%
A_{D}a^{D+1}}\int_{0}^{\infty }dx\,x^{D+1}\frac{K_{\nu -1}(x)}{I_{\nu -1}(x)}%
S_{2\nu }^{(\mu )}(xz/z_{2}),  \label{EMTb2}
\end{equation}%
is the part in the VEV\ induced in the region $z<z_{2}$ by a single boundary
at $z=z_{2}$ when the boundary at $z=z_{1}$ is absent. In (\ref{EMT}) and (%
\ref{EMTb2}) we have introduced the notations%
\begin{eqnarray}
S_{2\nu }^{(l)}(x) &=&\frac{1}{D}\left[ I_{\nu -1}^{2}(x)-I_{\nu }^{2}(x)%
\right] ,  \notag \\
S_{2\nu }^{(D)}(x) &=&I_{\nu }^{2}(x)-I_{\nu -1}^{2}(x)+\frac{2\nu -1}{x}%
I_{\nu }(x)I_{\nu -1}(x),  \label{S2nu}
\end{eqnarray}%
and
\begin{eqnarray}
F_{2\nu }^{(l)}(x,y) &=&\frac{1}{D}\left[ G_{\nu -1,\nu -1}^{2}(x,y)-G_{\nu
-1,\nu }^{2}(x,y)\right] ,  \notag \\
F_{2\nu }^{(D)}(x,y) &=&G_{\nu -1,\nu }^{2}(x,y)-G_{\nu -1,\nu -1}^{2}(x,y)-%
\frac{2\nu -1}{y}G_{\nu -1,\nu -1}(x,y)G_{\nu -1,\nu }(x,y),  \label{F2nu}
\end{eqnarray}%
with $l=0,\ldots ,D-1$. The last term in (\ref{EMT}) is induced by the right
boundary. It is finite for $z_{1}<z\leq z_{2}$ and diverges at $z=z_{1}$. It
can be seen that, for a massless field one gets (no summation over $\mu $) $%
\langle T_{\mu }^{\mu }\rangle
_{2}^{\mathrm{(b)}}=(z/a)^{D+1}\langle T_{\mu }^{\mu }\rangle
_{\mathrm{(M)},m=0}$, where $\langle T_{\mu }^{\mu }\rangle
_{\mathrm{(M)},m=0}$ is given by (\ref{EMTm0}) with $z_{1}=0$. As
we see, in the region $z<z_{2}$ the problem with a single boundary
in the AdS bulk is conformally related to the problem in the
Minkowski spacetime with two boundaries. This is a consequence of
the boundary condition we have imposed on the AdS boundary.

It can be checked that both the single-boundary induced part, $\langle
T_{\alpha }^{\mu }\rangle _{j}^{\mathrm{(b)}}$, and the second
boundary-induced part (last terms in (\ref{EMT}) and (\ref{EMTb2})) in the
VEV of the energy-momentum tensor obey the trace relation $T_{\mu }^{\mu }=m%
\bar{\psi}\psi $. In addition, the VEV obeys the covariant continuity
equation $T_{\alpha }^{\mu }{}_{;\mu }=0$ which, for the geometry under
consideration, reduces to the single equation
\begin{equation}
z^{D+2}\partial _{z}\left( T_{D}^{D}/z^{D+1}\right) +T_{\mu }^{\mu }=0.
\label{ContEq}
\end{equation}%
Using the trace relation and taking into account that the boundary-induced
part in the FC is negative everywhere, from (\ref{ContEq}) we conclude that
the boundary-induced part in the VEV of the normal stress is a monotonically
increasing function of $z$ for all points outside the boundaries.

Taking into account that $I_{\nu }(x)<I_{\nu -1}(x)$ and $K_{\nu }(x)>K_{\nu
-1}(x)$, from (\ref{EMTb1}) and (\ref{EMTb2}) we see that, for the geometry
of a single boundary at $z=z_{j}$, the boundary-induced part in the energy
density is negative everywhere: $\langle T_{0}^{0}\rangle _{j}^{\mathrm{(b)}%
}<0$. Using the properties of the modified Bessel functions, it can be seen
that $S_{j\nu }^{(D)}(x)>0$ for $z>z_{j}$ and $S_{j\nu }^{(D)}(x)<0$ for $%
z<z_{j}$. From here it follows that $\langle T_{D}^{D}\rangle _{j}^{\mathrm{%
(b)}}<0$ in the region $z>z_{j}$ and $\langle T_{D}^{D}\rangle _{j}^{\mathrm{%
(b)}}>0$ in the region $z<z_{j}$. Next, it can be checked that%
\begin{eqnarray}
G_{\nu ,\nu -1}(x,y) &>&-G_{\nu ,\nu }(x,y)>0,\;x<y,  \notag \\
G_{\nu -1,\nu }(x,y) &>&G_{\nu -1,\nu -1}(x,y)>0,\;x>y,  \label{RelG}
\end{eqnarray}%
From these relations we see that $F_{j\nu }^{(0)}(x,y)<0$ and the parts of
the energy density induced by the second plate (last terms in (\ref{EMT2})
and (\ref{EMT}) with $\mu =0$) are negative. Hence, for the geometry of two
boundaries the energy density is negative everywhere.

Now let us consider the asymptotics for the single boundary parts in the VEV
of the energy-momentum tensor at small and large distances from the
boundary. For the boundary at $z=z_{j}$, at large distances, $z\gg z_{j}$,
to leading order one has%
\begin{equation}
\langle T_{D}^{D}\rangle _{j}^{\mathrm{(b)}}=\frac{D\langle T_{0}^{0}\rangle
_{j}^{\mathrm{(b)}}}{D+2\nu }=-D\frac{Nma^{-D}(z_{j}/z)^{2\nu }}{2^{D+2\nu
+2}\pi ^{(D-1)/2}}\frac{\Gamma (D/2+\nu )\Gamma (D/2+2\nu )}{\nu \Gamma
^{2}(\nu )\Gamma (D/2+\nu +3/2)},  \label{EMTLarge}
\end{equation}%
with $z_{j}/z=e^{-(y-y_{j})/a}$. For the region $z<z_{j}$ with the condition
$z\ll z_{j}$, the leading order terms have the form:%
\begin{equation}
\langle T_{D}^{D}\rangle _{j}^{\mathrm{(b)}}\approx -\frac{D}{2\nu }\langle
T_{0}^{0}\rangle _{j}^{\mathrm{(b)}}\approx \frac{Na^{-D-1}(z/z_{j})^{D+2\nu
}}{2^{2\nu -1}A_{D}\nu \Gamma ^{2}(\nu )}\int_{0}^{\infty }dx\,x^{D+2\nu -1}%
\frac{K_{\nu -1}(x)}{I_{\nu -1}(x)}.  \label{EMTLarge1}
\end{equation}%
The relations between the energy density and the normal stress, given by (%
\ref{EMTLarge}) and (\ref{EMTLarge1}), can also be obtained by using the
continuity equation (\ref{ContEq}) for $\langle T_{\alpha }^{\mu }\rangle
_{j}^{\mathrm{(b)}}$ with $\langle T_{\mu }^{\mu }\rangle _{j}^{\mathrm{(b)}%
}=D\langle T_{0}^{0}\rangle _{j}^{\mathrm{(b)}}+\langle T_{D}^{D}\rangle
_{j}^{\mathrm{(b)}}$. As it is seen from (\ref{EMTLarge}) and (\ref%
{EMTLarge1}), at distances from the boundary larger than the AdS curvature
scale, the boundary-induced part in the VEV\ of the energy-momentum tensor
decays exponentially. For a scalar field with curvature coupling parameter $%
\xi $ and with Robin boundary condition at $z=z_{j}$, for the VEV of the
energy-momentum tensor one has \cite{Saha05}: $\langle T_{\mu }^{\mu
}\rangle _{j}^{\mathrm{(b)}}$ $\propto (z_{j}/z)^{2\nu _{\mathrm{sc}}}$ for $%
z\gg z_{j}$ and $\langle T_{\mu }^{\mu }\rangle _{j}^{\mathrm{(b)}}$ $%
\propto (z_{j}/z)^{D+2\nu _{\mathrm{sc}}}$ for $z\ll z_{j}$, where $\nu _{%
\mathrm{sc}}=\sqrt{D^{2}/4-D(D+1)\xi +m^{2}a^{2}}$. In particular, for a
conformally coupled scalar one has $\nu _{\mathrm{sc}}=\sqrt{m^{2}a^{2}+1/2}$%
, and the suppression of the VEVs is weaker than in the fermionic case with
the same value of the nonzero mass.

For points near the boundary, the dominant contributions to the integrals in
(\ref{EMTb1}) and (\ref{EMTb2}) come from large values of $x$. By using the
asymptotic expressions for the modified Bessel functions for large values of
the argument, we get the following leading behavior:%
\begin{equation}
\langle T_{0}^{0}\rangle _{j}^{(b)}\approx \frac{\left( D-1\right) a}{%
D\left( y-y_{j}\right) }\langle T_{D}^{D}\rangle _{j}^{(b)}\approx -\frac{%
Nm\Gamma ((D+1)/2)}{D(4\pi )^{(D+1)/2}|y-y_{j}|^{D}}.  \label{EMTnear}
\end{equation}%
For a boundary in the Minkowski spacetime the leading terms for the energy
density and stresses along directions parallel to the boundary coincide with
(\ref{EMTnear}), whereas the normal stress vanishes.

The Minkowskian limit of the formulas for the VEV of the energy-momentum
tensor is taken in a way similar to that we used for the case of FC. For $%
ma\gg 1$ and for a fixed value of $y$, to leading order we find (no
summation over $\mu $):%
\begin{equation}
\langle T_{\mu }^{\mu }\rangle \approx \langle T_{\mu }^{\mu }\rangle _{%
\mathrm{(M)}}=-\frac{N}{A_{D}}\int_{m}^{\infty }dx\,\frac{%
(x^{2}-m^{2})^{D/2-1}}{\frac{x+m}{x-m}e^{2x(y_{2}-y_{1})}+1}G_{\mathrm{(M)}%
}^{(\mu )}(x),  \label{EMTRD}
\end{equation}%
where%
\begin{equation}
G_{\mathrm{(M)}}^{(\mu )}(x)=\frac{x^{2}-m^{2}}{D}\left\{ 2+\frac{m}{x-m}%
\left[ e^{2x(y-y_{1})}+e^{2x(y_{2}-y)}\right] \right\} ,  \label{EMTM}
\end{equation}%
for $\mu =0,1,\ldots ,D-1$, and $G_{\mathrm{(M)}}^{(D)}(x)$ $=-2x^{2}$.
Here, $\langle T_{\mu }^{\mu }\rangle _{\mathrm{(M)}}$ is the VEV for the
geometry of two boundaries in the Minkowski bulk. The formula (\ref{EMTRD})
is a special case of the result given in \cite{Eliz11}. The fermion Casimir
energy for two parallel plates in 4-dimensional Minkowski spacetime has been
investigated in \cite{John75} and \cite{Mama80} for massless and massive
fields, respectively. The corresponding result for arbitrary number of
dimensions is generalized in \cite{Paol99}. The topological Casimir effect
and the VEV of the fermionic current for a massive fermionic field in a
spacetime with an arbitrary number of toroidally compact dimensions have
been considered in \cite{Bell09b}.

\begin{figure}[tbph]
\begin{center}
\epsfig{figure=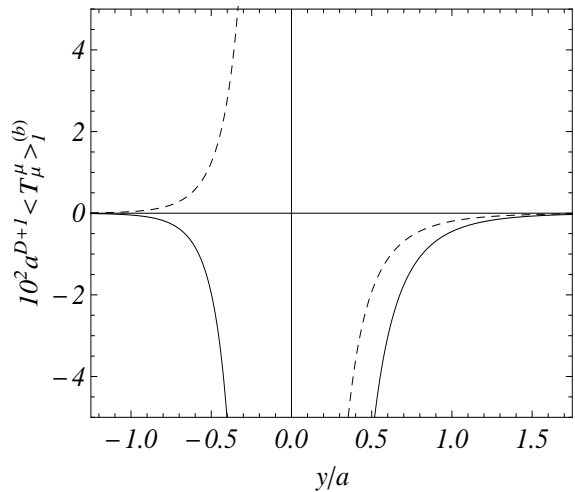,width=7.5cm,height=6.5cm}
\end{center}
\caption{Boundary-induced parts of the VEV of the energy density (full
curve) and of the normal stress (dashed curve), induced by a single
boundary, at $y=0$, as functions of $y/a$. Both plots are for $D=3$ and $%
ma=1 $.}
\label{fig1}
\end{figure}

In Fig.~\ref{fig1}, for the geometry of a single boundary, located at $y=0$,
we display the boundary-induced parts of the VEV of the energy density ($\mu
=0$, full curve) and of the normal stress ($\mu =D$, dashed curve), as
functions of $y/a$. The latter measures the distance from the boundary in
units of the AdS curvature radius. The graphs are plotted for a fermionic
field in 4-dimensional AdS spacetime ($D=3$) and for the mass we have taken $%
ma=1$.

\section{Interaction forces and the Casimir energy}

\label{sec:Force}

The force acting per unit surface of the boundary at $z=z_{j}$ (vacuum
effective pressure) can be obtained evaluating the normal stress at the
location of the boundary: $p^{(j)}=$ $-$ $\langle T_{D}^{D}\rangle
_{z=z_{j}} $. The boundary-free parts of the force acting from the left- and
from the right-hand sides of the boundary compensate and the resulting force
is determined by the boundary induced part. By using the decomposition for
the VEV of the energy-momentum tensor, in the region between the boundaries,
the effective pressure is obtained as%
\begin{equation}
p^{(j)}=p_{1}^{(j)}+p_{\mathrm{(int)}}^{(j)},  \label{pj}
\end{equation}%
where the first term on the rhs correspond to the situation when the second
boundary is absent and the second term is induced by the presence of the
second boundary. The latter can be termed as the interaction contribution.
For the first part, one has $p_{1}^{(j)}=-\langle T_{D}^{D}\rangle
_{j,z=z_{j}}$. In the regions $z\leq z_{1}$ and $z\geq z_{2}$, the single
boundary terms remain only: $p^{(1)}=p_{1}^{(1)}$ for $z<z_{1}$ and $%
p^{(2)}=p_{1}^{(2)}$ for $z>z_{2}$. Because of the surface divergences in
the single-boundary parts of the VEV, the term $p_{1}^{(j)}$ is divergent
and needs renormalization. The interaction part is finite for all nonzero
values of the distance between the boundaries and it is not affected by the
renormalization procedure.

The interaction parts of the effective pressure are obtained from the last
terms in (\ref{EMT2}) and (\ref{EMT}), by setting in the expressions for the
$_{D}^{D}$-components $z=z_{1}$ and $z=z_{2}$, respectively. Using the
Wronskian relation for the modified Bessel functions, one finds%
\begin{equation}
p_{\mathrm{(int)}}^{(j)}=-\frac{N(z_{j}/z_{1})^{D}}{A_{D}a^{D+1}}%
\int_{0}^{\infty }dx\,x^{D-1}\Omega _{\nu }^{(j)}(x,\eta x).  \label{pjint}
\end{equation}%
As we see, the corresponding effective pressures are always negative and,
hence, the interaction forces are attractive. Using now the relations%
\begin{equation}
\Omega _{\nu }^{(j)}(xz_{1},xz_{2})=(-1)^{j-1}z_{j}\partial _{z_{j}}\ln
\left[ 1+\frac{I_{\nu }(xz_{1})K_{\nu -1}(xz_{2})}{K_{\nu }(xz_{1})I_{\nu
-1}(xz_{2})}\right] ,  \label{DerRel}
\end{equation}%
the interaction parts can be expressed in a more convenient way as%
\begin{equation}
p_{\mathrm{(int)}}^{(j)}=\frac{N(z_{j}{}/a)^{D+1}}{(-1)^{j}A_{D}}\partial
_{z_{j}}\int_{0}^{\infty }dx\,x^{D-1}\ln \left[ 1+\frac{I_{\nu
}(xz_{1})K_{\nu -1}(xz_{2})}{K_{\nu }(xz_{1})I_{\nu -1}(xz_{2})}\right] .
\label{pjint2}
\end{equation}%
Note that the forces acting on the left and on the right boundaries are
different, in general. This property is related to the nonzero extrinsic
curvature tensor for the the boundary geometry under consideration.

At small distances between the boundaries, $\eta -1\ll 1$, the dominant
contribution comes from large values of $x$ and, to leading order, we get%
\begin{equation}
p_{\mathrm{(int)}}^{(j)}\approx -\frac{ND\Gamma (\left( D+1\right) /2)}{%
(4\pi )^{(D+1)/2}(y_{2}-y_{1})^{D+1}}.  \label{pjintsmall}
\end{equation}%
At large distances one has $\eta \gg 1$, and the leading terms in the
asymptotic expansions are given by the expressions%
\begin{eqnarray}
p_{\mathrm{(int)}}^{(1)} &\approx &-\frac{2^{2-2\nu }Na^{-D-1}}{A_{D}\Gamma
^{2}(\nu )\eta ^{D+2\nu }}\int_{0}^{\infty }dx\,x^{D+2\nu -1}\frac{K_{\nu
-1}(x)}{I_{\nu -1}(x)},  \notag \\
p_{\mathrm{(int)}}^{(2)} &\approx &-\frac{2^{1-2\nu }Na^{-D-1}}{A_{D}\nu
\Gamma ^{2}(\nu )\eta ^{2\nu }}\int_{0}^{\infty }dx\,\frac{x^{D+2\nu -1}}{%
I_{\nu -1}^{2}(x)}.  \label{pjfar}
\end{eqnarray}

Now, let us consider the Minkowskian limit corresponding to $ma\gg 1$ for a
fixed value of $y$. In this limit one has $\eta \approx 1+(y_{2}-y_{1})/a$.
Introducing in (\ref{pjint}) a new integration variable, $u=x/\nu $, we use
now the uniform asymptotic expansions for the modified Bessel functions for
large values of the order. After some transformations, to leading order we
have%
\begin{equation}
p_{\mathrm{(int)}}^{(j)}\approx p_{\mathrm{(M)}}=-\frac{2N}{A_{D}}%
m^{D+1}\int_{1}^{\infty }dx\,\frac{x^{2}(x^{2}-1)^{D/2-1}}{\frac{x+1}{x-1}%
e^{2xm(y_{2}-y_{1})}+1}.  \label{pjMink}
\end{equation}%
Of course, in the Minkowskian limit the forces are the same for the
boundaries at $y=y_{1}$ and $y=y_{2}$. Note that, for the geometry of a
single boundary, in the Minowskian limit the stresses on the left- and
right-hand sides are the same by the symmetry of the problem. As a result,
the corresponding net force vanishes and only the interaction part remains.

\begin{figure}[tbph]
\begin{center}
\epsfig{figure=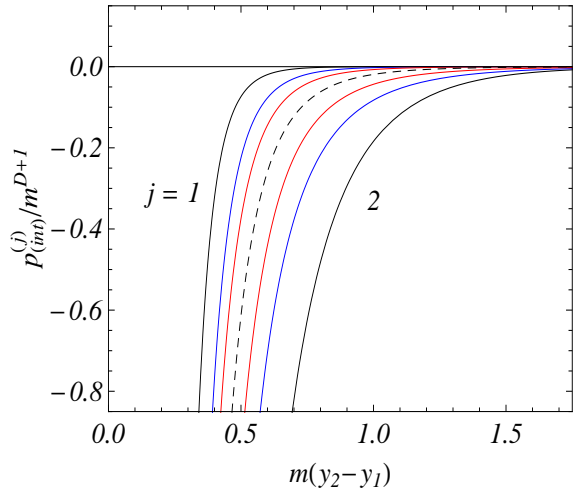,width=7.5cm,height=6.5cm}
\end{center}
\caption{Interaction forces per unit surface as functions of the separation
between the boundaries for the fermionic field in a 4-dimensional AdS bulk.
The dashed curve corresponds to the force for boundaries in the Minkowski
bulk. The plots correspond to $ma=0.5$ (black), $ma=1$ (blue) and $ma=2$
(red), respectively.}
\label{fig2}
\end{figure}

In Fig.~\ref{fig2}, we depict the interaction forces for the model with $D=3$
as functions of the separation between boundaries as measured in units of
the Compton wavelength of the fermionic particle. The dashed curve is the
corresponding force for boundaries in Minkowski spacetime. The plots are for
$ma=0.5$ (black), $ma=1$ (blue) and $ma=2$ (red). The curves on the left
(right) of the dashed curve are for $j=1$ ($j=2$). As we see, with
increasing $a$ the forces tend to the corresponding result for the Minkowski
bulk.

Now we consider the total vacuum energy in the region between the
boundaries. The formal expression for this energy is obtained by integration
of the energy density given by (\ref{EMT1}), with $\mu =0$. Making use of
standard formula for the integrals involving the square of a cylinder
function, we get%
\begin{eqnarray}
E_{z_{1}\leq z\leq z_{2}} &=&\int_{z_{1}}^{z_{2}}dz\,\sqrt{|g|}\langle
T_{0}^{0}\rangle  \notag \\
&=&-\frac{N}{2}\int \frac{d\mathbf{k}}{(2\pi )^{D-1}}\,\mathbf{\,}%
\sum_{n=1}^{\infty }\sqrt{\gamma _{\nu ,n}^{2}/z_{1}^{2}+k^{2}}.  \label{E}
\end{eqnarray}%
This formula expresses the vacuum energy as a sum of ground state energies
for elementary oscillators. Obviously, expression (\ref{E}) is divergent and
regularization, with the subsequent renormalization, is necessary. Here we
follow the zeta function approach (for the application of the zeta function
techniques to the calculations of the Casimir energy see \cite{Kirs01} and
references therein). Instead of (\ref{E}), we consider the related zeta
function%
\begin{equation}
\zeta (s)=-\mu _{0}^{s+1}\frac{N}{2}\int \frac{d\mathbf{k}}{(2\pi )^{D-1}}%
\mathbf{\,}\sum_{n=1}^{\infty }\left( \gamma _{\nu
,n}^{2}/z_{1}^{2}+k^{2}\right) ^{-s/2},  \label{zeta}
\end{equation}%
where the constant $\mu _{0}$ with dimensions of mass is introduced by
dimensional reasons (it is the regularization parameter). The expression on
the rhs of (\ref{zeta}) is finite for $\mathrm{Re}s>D$ and for the
evaluation of the vacuum energy we need its analytic continuation at $s=-1$:
$E_{z_{1}\leq z\leq z_{2}}=\zeta (s)|_{s=-1}$.

Evaluating the integral over $\mathbf{k}$, the zeta function can be
expressed in the form%
\begin{equation}
\zeta (s)=-\frac{N\Gamma ((s+1-D)/2)\mu _{0}^{s+1}}{2(4\pi )^{(D-1)/2}\Gamma
(s/2)z_{1}^{D-1-s}}\zeta _{1}\left( s+1-D\right) \mathbf{\,},  \label{zeta1}
\end{equation}%
where we have introduced the partial zeta function%
\begin{equation}
\zeta _{1}\left( s\right) =\sum_{n=1}^{\infty }\gamma _{\nu ,n}^{-s},
\label{zetap}
\end{equation}%
directly related to the eigenvalues $\gamma _{\nu ,n}$. We need the analytic
continuation of the function (\ref{zetap}) to a neighborhood of $s=-D$. This
procedure is standard in the theory of the Casimir effect and we only give
the main steps.

From Cauchy's residue formula, the integral representation follows%
\begin{equation}
\zeta _{1}\left( s\right) =\int_{C}\frac{du}{2\pi i}\,\,u^{-s}\partial
_{u}\ln \left[ ug_{\nu ,\nu -1}(u,u\eta )\right] ,  \label{zetanint1}
\end{equation}%
where $C$ is a closed, counterclockwise contour on the complex $z$ plane
enclosing all zeros $\gamma _{\nu ,n}$. We take the contour made of a large
semicircle (with radius tending to infinity) centered at the origin and
placed to its right, plus a straight part overlapping the imaginary axis and
avoiding the origin by a small semicircle $C_{\rho }$ on the right
half-plane with radius $\rho $. For small $\rho $, one has%
\begin{equation}
\int_{C_{\rho }}\frac{du}{2\pi i}\,u^{-s}\partial _{u}\ln \left[ ug_{\nu
,\nu -1}(u,u\eta )\right] =B\frac{\rho ^{2-s}}{2-s}\sin (\pi s/2),
\label{Cro}
\end{equation}%
being $B$ a constant independent of $s$. Denoting the upper and lower halves
of the contour $C$ by $C_{1}$ and $C_{2}$, respectively, the integral can be
cast in the form%
\begin{eqnarray}
\zeta _{1}\left( s\right) &=&\int_{C}\frac{dz}{2\pi i}\,u^{-s}\partial
_{u}\ln \left[ u^{1-\nu }J_{\nu -1}(u\eta )\right]  \notag \\
&&+\sum_{l=1,2}\int_{C_{l}}\frac{du}{2\pi i}\,\,u^{-s}\partial _{u}\ln \left[
u^{\nu }H_{\nu }^{(\alpha )}(u)\right]  \notag \\
&&+\sum_{l=1,2}\int_{C_{l}}\frac{du}{2\pi i}\,\,u^{-s}\partial _{u}\ln \left[
1-\frac{J_{\nu }(u)H_{\nu -1}^{(\alpha )}(u\eta )}{H_{\nu }^{(\alpha
)}(u)J_{\nu -1}(u\eta z)}\right] .  \label{zetanint2}
\end{eqnarray}%
After parameterizing the integrals over the imaginary axis and substituting
into (\ref{zeta1}), we arrive at the following expression for the zeta
function%
\begin{eqnarray}
\zeta (s) &=&-\frac{(4\pi )^{(1-D)/2}N\mu _{0}^{s+1}z_{1}^{1+s-D}}{2\Gamma
(s/2)\Gamma ((D+1-s)/2)}\left\{ \frac{\pi B\rho ^{D+1-s}}{D+1-s}\right.
\mathbf{\,}  \notag \\
&&+\int_{\rho }^{\infty }dx\,x^{D-1-s}\partial _{x}\Bigg[\ln \left( x^{1-\nu
}I_{\nu -1}(x\eta )\right)  \notag \\
&&\left. +\ln \left( x^{\nu }K_{\nu }(x)\right) +\ln \left( 1+\frac{I_{\nu
}(x)K_{\nu -1}(x\eta )}{K(x)I_{\nu -1}(x\eta )}\right) \Bigg]\right\} ,
\label{zeta3}
\end{eqnarray}%
where we have used that $\Gamma (y)\sin (\pi y)=\pi /\Gamma (1-y)$. The term
with the factor $B$ vanishes in the limit $\rho \rightarrow 0$ at the
physical point $s=-1$, while the last term is finite at this point. The
first (second) term in the square brackets in (\ref{zeta3}) corresponds to
the vacuum energy in the region $z\leq z_{2}$ ($z\geq z_{1}$) for the
geometry of a single boundary at $z=z_{2}$ ($z=z_{1}$) and will be denoted
as $E_{1,z\leq z_{2}}^{(2)}$ ($E_{1,z\geq z_{1}}^{(1)}$). Adding also the
vacuum energies from the regions $z\leq z_{1}$ and $z\geq z_{2}$ (denoted as
$E_{1,z\leq z_{1}}^{(1)}$ and $E_{1,z\geq z_{2}}^{(2)}$), for the total
vacuum energy we finally get%
\begin{equation}
E=E_{1,z\leq z_{1}}^{(1)}+E_{1,z\geq z_{2}}^{(2)}+E_{z_{1}\leq z\leq
z_{2}}=\sum_{j=1,2}E_{1}^{(j)}+\Delta E.  \label{Etot}
\end{equation}%
Here, $E_{1}^{(j)}$ is the energy for the geometry of a single boundary at $%
z=z_{j}$, and the interference part $\Delta E$ is given by the last term in (%
\ref{zeta3}) with $s=-1$. After integration by parts, one gets for the latter%
\begin{equation}
\Delta E=-\frac{N}{A_{D}}\int_{0}^{\infty }dx\,x^{D-1}\ln \left[ 1+\frac{%
I_{\nu }(xz_{1})K_{\nu -1}(xz_{2})}{K(xz_{1})I_{\nu -1}(xz_{2})}\right] .
\label{DeltaE}
\end{equation}%
The renormalization procedure to be carried out for the divergences of the
single boundary parts in (\ref{Etot}) is similar to those previously
discussed within the framework of braneworld scenarios (see, e.g., \cite%
{Gold00,Flac01}). Note that the dependence of $E_{1}^{(j)}$ on $z_{j}$ is in
the form $z_{j}^{-D}$ and this dictates the form of the counterterms located
on the boundaries. By finite renormalizations, the single boundary terms in (%
\ref{Etot}) can be absorbed into the counterterms. The renormalized vacuum
energy has the form $E=\sum_{j=1,2}c_{j}z_{j}^{-D}+\Delta E$, with
renormalized coefficients $c_{j}$. Comparing (\ref{DeltaE}) with (\ref%
{pjint2}), we find the relation%
\begin{equation}
p_{\mathrm{(int)}}^{(j)}=-(-1)^{j}(z_{j}{}/a)^{D+1}\partial _{z_{j}}\Delta E,
\label{RelpE}
\end{equation}%
between the interaction forces and the interference part of the Casimir
energy.

Here we have considered the fermionic Casimir effect with bag boundary
conditions on the background of AdS spacetime. In a similar way the
fermionic Casimir densities can be evaluated for the Randall-Sundrum
braneworld model. This model is formulated on 5-dimensional AdS spacetime,
thus with a single extra dimension. The fifth dimension $y$ is compactified
on an orbifold $S^{1}/Z_{2}$ of length $L$, with $-L\leq y\leq L$. The
corresponding line element has the form (\ref{ds2deSit}) with the warp
factor $e^{-2|y|/a}$. Two 3-branes are located at the orbifold fixed points,
$y=0$ and $y=L$. In terms of the conformal radial coordinate $z$, for the
branes one has $z=a$ and $z=z_{L}=ae^{L/a}$. In the Randall-Sundrum model,
depending on the parity of the spinor field under a chiral transformation,
two types of boundary conditions arise on the branes. For these boundary
conditions the eigenvalues of the quantum number $\lambda $ are roots of the
equation%
\begin{equation}
g_{\nu -s,\nu -s}(\lambda a,\lambda z_{L})=0,  \label{RSbc}
\end{equation}%
with $s=0,1$ for even and odd fields, respectively. The summation formula
for series over these eigenvalues is obtained from the general results of
\cite{Saha01,SahaBook}. Calculations are actually the same as those we have
described above for the case of bag boundary conditions. In the
normalization condition for the mode functions the integration over $y$ goes
over the region $(-L,L)$. As a result the normalization coefficient will
have an additional factor 1/2, as compared with the case where the problem
is formulated on the interval $(0,L)$.

The FC and the VEV of the energy-momentum tensor for a fermionic field in
the Randall-Sundrum model are obtained from the formulas given in Sects.~\ref%
{sec:FC} and \ref{sec:EMT} by changing the order of the appropriately
modified Bessel function from $\nu -1$ to $\nu $ for even fields and from $%
\nu $ to $\nu -1$ for odd fields and by adding an extra factor of 1/2. For
example, in the case of even fields, to Eq.~(\ref{FCb}) we need only add the
factor 1/2, whereas to Eq.~(\ref{FCbLeft}) we have to add the factor 1/2 and
make also the replacements $K_{\nu -1}(x)\rightarrow -K_{\nu }(x)$ and $%
I_{\nu -1}(x)\rightarrow I_{\nu }(x)$. In the case of odd fields the
situation is just opposite: in (\ref{FCb}) we add the factor 1/2 and replace
$K_{\nu }(x)\rightarrow -K_{\nu -1}(x)$ and $I_{\nu }(x)\rightarrow I_{\nu
-1}(x)$, while in (\ref{FCbLeft}) we only add the factor 1/2. Note that when
evaluating the vacuum energy in the braneworld model the integration goes
over the region $(-L,L)$ and there is no need to add the factor 1/2 in the
corresponding expressions.

Through the above mentioned replacements of the modified Bessel functions,
from (\ref{DeltaE}) we readily obtain the corresponding vacuum energies for
even and odd fields in the Randall-Sundrum model. For $D=4$ the
corresponding formulas were obtained in \cite{Flac01} (note that in this
reference the effective Lagrangian is considered per fermionic degree of
freedom, which corresponds to $-\Delta E/N$). The VEV of the energy-momentum
tensor for a bulk Dirac spinor in the Randall-Sundrum model has been
considered in \cite{Shao10}. In this reference, for the case of a massive
field, a general formula is given for the unrenormalized VEV only. To
compare, in our approach, based on the generalized Abel-Plana formula, the
pure AdS parts in the VEVs are extracted explicitly and, for the points away
from the branes, the renormalization procedure is the same as for the
boundary-free parts. In addition, the boundary induced parts are presented
in terms of exponentially convergent integrals, which are very well suited
for numerical calculations.

\section{Conclusions}

\label{sec:Conc}

In this paper we have investigated the fermionic condensate and the VEV of
the energy-momentum tensor for a massive fermionic field in AdS spacetime in
the presence of two boundaries on which the field obeys bag boundary
conditions. For the evaluation of the VEVs we have employed the mode
summation technique. In the region between the boundaries, a complete set of
positive- and negative-energy mode functions is given by (\ref{psi+}) and (%
\ref{psi-}), respectively, where the eigenvalues of the radial quantum
number $\lambda $ are determined from the boundary conditions and they are
solutions of the equation (\ref{lambModes}). The mode sums for the FC and
the energy-momentum tensor contain series over these eigenvalues. For the
summation of the series we have used the generalized Abel-Plana formula (\ref%
{SumForm}), which allowed to separate the VEVs into single boundary and
second boundary-induced parts. In this representation, explicit knowledge of
the eigenvalues of $\lambda $ is not necessary. The VEVs for the geometry of
a single boundary are further decomposed into boundary-free and
boundary-induced parts. As a result, in the region between the boundaries,
the VEVs can be expressed in two equivalent ways, respectively given by
Eqs.~(\ref{FC4}) and (\ref{FC5}), for the FC, and by Eqs.~(\ref{EMT2}) and (%
\ref{EMT}), for the energy-momentum tensor. With these representations, and
for points away from the boundaries, the boundary induced part is finite and
renormalization is required for the boundary-free part, only.

For the geometry of a single boundary located at $z=z_{j}$, the
boundary-induced contribution to the FC is given by (\ref{FCb}), in the
region $z>z_{j}$, and by (\ref{FCbLeft}), in the region $z<z_{j}$. This
contribution is negative for both regions and it is not symmetric with
respect to the boundary. Such fact is related to the existence of a nonzero
extrinsic curvature tensor for the boundary $z=z_{j}$ in the AdS bulk. At
large distances from the boundary, as compared with the value of the AdS
curvature radius, the boundary-induced parts are exponentially suppressed by
the factors $e^{-2\nu (y-y_{j})/a}$, in the region $y>y_{j}$, and $%
e^{-(D+2\nu )(y-y_{j})/a}$, in the region $y<y_{j}$. In particular, the
boundary-induced part vanishes on the AdS boundary. For points near the
boundary at $y=y_{j}$, the leading term in the asymptotic expansion of the
FC is given by (\ref{FCnear}) and it does coincide with the corresponding
expression for the boundary in Minkowski spacetime. For the geometry of two
boundaries, the FC in the region between them can be expressed in the form (%
\ref{FCInt}), where the interference term is finite everywhere, including
the points on the boundaries. For large separation of the boundaries, as
compared with the curvature radius of the background spacetime, the
interference part is exponentially suppressed. The boundary-induced part of
the FC in the region between the two boundaries is negative.

The boundary-induced contributions in the VEV of the energy-momentum tensor
for a single boundary are given by (\ref{EMTb1}) and (\ref{EMTb2}), for the
regions on the right and on the left of the boundary, respectively. The
corresponding vacuum energy is negative, whereas the normal stress is
negative on the right domain and positive on the left one. At large
distances from a single boundary located at $y=y_{j}$, the boundary-induced
terms decay as $e^{-2\nu (y-y_{j})/a}$, for $y>y_{j}$, and as $e^{-(D+2\nu
)(y-y_{j})/a}$, for $y<y_{j}$. For points near the boundary the
corresponding asymptotic behavior is given by (\ref{EMTnear}).

The forces acting on the boundaries and the Casimir energy were considered
in Sect.~\ref{sec:Force}. The force acting on the boundary at $z=z_{j}$ is
decomposed as (\ref{pj}) where the first term on the rhs is the force for a
single boundary (when the second one is absent) while the second term is
induced by the presence of the other boundary. The interaction part of the
force is attractive and it can be expressed in the form (\ref{pjint}) or,
equivalently, as in (\ref{pjint2}). The forces acting on the left and on the
right boundaries are different from each other. For small separations of the
boundaries, as compared with the AdS curvature radius, to leading order we
recover the result for the Minkowski bulk with boundaries. For large
separations, the asymptotic expressions for the force are given by (\ref%
{pjfar}) and the interaction force is again exponentially suppressed. We
have also checked with care the transition to the Minkowskian limit
corresponding to $a\rightarrow \infty $. For the evaluation of the total
vacuum energy in the region between the boundaries we have employed zeta
function techniques. As the corresponding scheme is well described in the
literature on the Casimir effect, we have here sketched the main steps only.
For the total vacuum energy, including the contributions coming from the
regions $z\leq z_{1}$ and $z\geq z_{2}$, one has (\ref{Etot}) with the
interference part being given by (\ref{DeltaE}). In the second part of Sect.~%
\ref{sec:Force} we have described in detail how from our results the
corresponding formulae for fermionic Casimir densities in
Randall-Sundrum-type braneworld scenarios immediately follow.

\section*{Acknowledgments}

A.A.S. was supported by the ESF Programme \textquotedblleft New Trends and
Applications of the Casimir Effect". E.E. and S.D.O. were partially funded
by MICINN (Spain), project FIS2010-15640, by the CPAN Consolider Ingenio
Project, and by AGAUR (Catalonia), project 2009SGR-994.

\end{document}